\documentclass[3p]{elsarticle}
\usepackage{amssymb,epsfig,amsmath,amsfonts,bm,cases}
\usepackage[utf8]{inputenc}

\makeatletter
\def\ps@pprintTitle{%
  \let\@oddhead\@empty
  \let\@evenhead\@empty
  \let\@oddfoot\@empty
  \let\@evenfoot\@oddfoot
}
\makeatother

\begin{document}

\title{Midpoint geometric integrators for inertial magnetization dynamics}

\author[1]{M. d'Aquino\corref{cor1}}%
\ead{mdaquino@unina.it}
\author[1]{S. Perna}
\ead{salvatore.perna@unina.it}
\author[1]{C. Serpico}
\ead{serpico@unina.it}
\cortext[cor1]{Corresponding author}

\affiliation[1]{organization={Department of Electrical Engineering and Information Technology, University of Naples Federico II},
addressline={Via Claudio 21},
postcode={I-80125},
city={Naples},
country={Italy}}

\begin{abstract}
We consider the numerical solution of the inertial version of Landau-Lifshitz-Gilbert equation (iLLG), which describes high-frequency nutation on top of magnetization precession due to angular momentum relaxation. The iLLG equation defines a higher-order nonlinear dynamical system with very different nature compared to the classical LLG equation, requiring twice as many degrees of freedom for space-time discretization. It exhibits essential conservation properties, namely magnetization amplitude preservation, magnetization projection conservation, and a balance equation for generalized free energy, leading to a Lyapunov structure (i.e. the free energy is a  decreasing function of time) when the external magnetic field is constant in time. We propose two second-order numerical schemes for integrating the iLLG dynamics over time, both based on implicit midpoint rule. The first scheme unconditionally preserves all the conservation properties, making it the preferred choice for simulating inertial magnetization dynamics. However, it implies doubling the number of unknowns, necessitating significant changes in numerical micromagnetic codes and increasing computational costs especially for spatially inhomogeneous dynamics simulations. To address this issue, we present a second time-stepping method that retains the same computational cost as the implicit midpoint rule for classical LLG dynamics while unconditionally preserving magnetization amplitude and projection. Special quasi-Newton techniques are developed for solving the nonlinear system of equations required at each time step due to the implicit nature of both time-steppings. The numerical schemes are validated on analytical solution for macrospin terahertz frequency response and the effectiveness of the second scheme is demonstrated with full micromagnetic simulation of inertial spin waves propagation in a magnetic thin-film.
    \end{abstract}

\begin{keyword} magnetic inertia \sep terahertz spin nutation \sep micromagnetic simulations \sep inertial Landau-Lifshitz-Gilbert (iLLG) equation \sep implicit midpoint rule \sep numerical methods. 
\end{keyword}
\maketitle

\section{Introduction}

The study of ultra-fast magnetization processes has become increasingly important in recent years, particularly for its potential applications to future generations of nanomagnetic and spintronic devices \cite{dieny2020opportunities}. Since the pioneering experiment by Beaurepaire et al. \cite{beaurepaire1996ultrafast} that revealed subpicosecond spin dynamics, the investigation of ultra-fast magnetization processes has attracted the attention of many research groups, leading to a considerable body of research \cite{koopmans2000ultrafast, stamm2007femtosecond, stanciu2007all, kimel2009inertia, kirilyuk2010ultrafast, lambert2014all, dornes2019ultrafast, hudl2019nonlinear}.

Recently, there have been exciting experimental developments in the direct detection of spin nutation in ferromagnets in the terahertz range \cite{neeraj2021inertial, Unikandanunni_PRL_2022}. This has confirmed the presence of inertial effects in magnetization dynamics, which were theoretically predicted several years ago \cite{ciornei2011magnetization, olive2012beyond, Mondal2017}. Nutation-like magnetization motions in nanomagnets occurring at gigahertz frequencies under the action of time-harmonic applied external magnetic fields were also studied theoretically in past decades within the classical precessional dynamics\cite{serpico_quasiperiodic_2004}.

From a technological perspective, the observation of terahertz spin nutation opens up new possibilities for exploiting novel ultra-fast regimes. For instance, it may be possible to use strong picosecond field pulses to drive ballistic magnetization switching into the inertial regime \cite{bauer2000switching, bertotti_geometrical_2003, daquino_numerical_2004, devolder2006precessional, neeraj2022inertial, Winter2022}. This has important implications for the development of ultra-fast magnetic devices, and it also has fundamental implications for the physics of magnetism.

From a theoretical point of view, inertial magnetization dynamics can be described by augmenting the classical Landau-Lifshitz-Gilbert (LLG) precessional dynamics with a torque term modeling intrinsic angular momentum relaxation \cite{ciornei2011magnetization,olive2012beyond}. 
This approach has been successful in explaining the observed high-frequency spin nutation in uniformly-magnetized ferromagnetic samples \cite{neeraj2021inertial}, for which magnetization dynamics is governed by the following inertial version of the Landau-Lifshitz-Gilbert equation\cite{ciornei2011magnetization,olive2012beyond}:
\begin{equation}\label{eq:iLLG macrospin physical}
    \frac{d\bm M}{dt}=-\gamma \bm M\times \left(\bm H_\mathrm{eff}-\frac{\alpha}{\gamma M_s}\frac{d\bm M}{dt}-{\tau^2}\frac{d^2\bm M}{dt^2}\right) \quad,
\end{equation}
where $\bm M(t)$ is the magnetization vector field ($M_s$ is the saturation magnetization of the material), $\bm H_\mathrm{eff}$ is the magnetic effective field, $\alpha$ is the Gilbert damping, $\gamma$ is the absolute value of the gyromagnetic ratio and $\tau$ defines the time scale of inertial magnetic phenomena. 

However, when spatial changes of magnetization do occur in magnetic systems of nano- and micro-scale, the description of spatially-inhomogeneous ultra-fast magnetization dynamics occurring at sub-picosecond time scales becomes a challenging problem that requires appropriate extension of eq.\eqref{eq:iLLG macrospin physical} to take into account space-varying vector fields in the region $\Omega$ occupied by the ferromagnetic body. This extension leads to the formulation of a novel equation where formally the total derivatives with respect to time become partial and the effective field is given by the variational derivative of the Gibbs-Landau free energy functional\cite{brown_micromagnetics_1963}, resulting in the following:
\begin{equation}\label{eq:iLLG physical}
    \frac{\partial\bm M}{\partial t}=-\gamma \bm M\times \left(\bm H_\mathrm{eff}-\frac{\alpha}{\gamma M_s}\frac{\partial\bm M}{\partial t}-{\tau^2}\frac{\partial^2\bm M}{\partial t^2}\right) \quad,
\end{equation}
where generally the natural (homogeneous Neumann) boundary conditions $\partial \bm M/\partial \bm n=0$ are inherited by the classical LLG when no surface anisotropy is present at the body surface $\partial\Omega$. Equation \eqref{eq:iLLG physical} reduces to the purely precessional classical LLG equation when no inertia is considered (i.e. $\tau=0$). Nevertheless, despite this apparent similarity, eq.\eqref{eq:iLLG physical} has profoundly different nature in that it has hyperbolic (wave-like) character (instead of parabolic as the classical LLG equation) and admits the possibility of travelling solutions (spin waves) with finite propagation speed\cite{daquino2023micromagnetic}. For this reason, the iLLG dynamics deserves a dedicated investigation in his own rights.   

In this respect, based on equations \eqref{eq:iLLG macrospin physical},\eqref{eq:iLLG physical}, a number of theoretical studies have been proposed in the latest years to characterize terahertz spin nutation\cite{Kikuchi2015,Giordano2020,Makhfudz2020nutation,lomonosov2021anatomy,cherkasskii2021dispersion,mondal2022inertial,Titov2022,Gareeva2023}. Most of these interesting studies rely on analytical approaches valid in idealized situations such as, for instance, analysis of magnetization oscillations in single-domain particles (macrospin) or small-amplitude spin waves propagation in infinite media.
Very recently, the possibility to observe propagation of ultra-short inertial spin waves in confined ferromagnetic thin-films driven by ac terahertz fields has been also theoretically demonstrated\cite{daquino2023micromagnetic}. These waves exhibit behavior that deviates significantly from classical exchange spin-waves and can propagate at a finite speed up to a limit of several thousands meters per second, which is comparable with the velocity of surface acoustic waves.

While such phenomena occurring in confined micromagnetic systems mainly involve magnetization oscillations around equilibria and can be investigated by analyzing the inertial LLG dynamics in the linear regime, no such possibility exists when far from equilibrium dynamics such as nonlinear oscillations\cite{Wigen1994}, magnetization switching\cite{neeraj2022inertial} or even chaos\cite{Montoya2019} are considered.

In these situations where no analytical techniques can be applied, one has to resort to numerical simulation.
In this respect, after the experimental evidence of the terahertz spin nutation\cite{neeraj2021inertial}, the study of inertial effects in magnetization dynamics is rapidly becoming an emergent field of research and, consequently, the need for accurate and efficient computational techniques that exploit the intrinsic properties of the nutation dynamics beyond off-the-shelf time-stepping schemes is growing fast, too. Nonetheless, at the present moment, very few works\cite{Ruggeri2022,Li2023} address ad-hoc numerical techniques for the inertial magnetization dynamics.

In this paper, after illustrating the general qualitative conservation properties of the continuous inertial magnetization dynamics, we propose suitable time-integration schemes based on the implicit
midpoint rule technique\cite{daquino_numerical_2005} for the numerical solution of the inertial LLG (iLLG) equation and their relevant properties are discussed. The midpoint rule is an unconditionally stable second order accurate scheme which preserves the fundamental geometrical properties of the classical LLG dynamics\cite{dAquino2005geometrical}. The first  time-stepping proposed here is shown to preserve all relevant conservation properties of the iLLG dynamics unconditionally, i.e. regardless of the time step amplitude.
Despite these remarkable properties, we show that, in general, the numerical integration of inertial magnetization dynamics must address the issue of the higher order of the dynamical system that it describes, which implies dramatic changes of micromagnetic codes and results anyway in at least doubling the computational cost of the numerical scheme as compared to  classical LLG dynamics. This has a huge impact when micromagnetic simulations with full spatial discretization on hundred thousands (or more) computational cells have to be performed, such as in the case of (sub)micron-sized magnetic systems. For this reason, we develop an additional efficient implementation of the
midpoint rule technique for iLLG dynamics, based on suitable multistep method for the inertial term, which can be built on the top of that associated with classical LLG dynamics and, therefore, retaining a computational cost with the same order of magnitude. The proposed techniques are first validated by computing the frequency response of a magnetic thin-film modeled as single spin (macrospin) magnetized along the easy direction and subject to out-of-plane ac field, and comparing the results with the analytical solution. Then, full micromagnetic simulations of inertial spin wave propagation in a ferromagnetic nanodot are performed in order to demonstrate the accuracy and effectiveness of the second proposed time-stepping in reproducing spatially-inhomogeneous ultra-fast spin nutation dynamics.

\section{Inertial magnetization dynamics and qualitative properties}

The starting point of the discussion is the inertial Landau-Lifshitz-Gilbert (iLLG) equation \eqref{eq:iLLG physical}, expressed in dimensionless form\cite{ciornei2011magnetization,daquino2023micromagnetic}:
\begin{equation}\label{eq:iLLG}
    \frac{\partial\bm m}{\partial t}=-\bm m \times \left(\bm h_\mathrm{eff}- \alpha\frac{\partial\bm m}{\partial t} - \xi \frac{\partial^2\bm m}{\partial t^2}\right) \,\,,
\end{equation}
where  $\bm m(\bm r,t)$ is the magnetization unit-vector (normalized by the saturation magnetization $M_s$) at each location $\bm r\in \Omega$ ($\Omega$ is the region occupied by the magnetic body), time is measured in units of $(\gamma M_s)^{-1}$ (corresponding to 5.7 ps for $\gamma=2.21\times 10^5 A^{-1}s^{-1} m$ and $\mu_0 M_s=1$T), $\alpha$ is the Gilbert (dimensionless and positive, typically in the order $\sim 10^{-3}\div 10^{-2}$) damping parameter, the parameter $\xi$ measures the strength of inertial effects in magnetization dynamics. It is worthwhile noting (see eq.\eqref{eq:iLLG physical}) that the dimensionless quantity $\xi$ can be expressed as $\xi=(\gamma M_s \tau)^2$ where $\tau$ determines the physical time-scale of magnetic inertia, for which previous works\cite{ciornei2011magnetization, neeraj2021inertial,neeraj2022inertial} assessed its order of magnitude as fractions of picosecond (this implies that typically $\xi\sim 10^{-2})$. Thus, the inertial effects in magnetization dynamics are governed by a quantity with the same smallness as usual Gilbert damping $\alpha\sim 10^{-2}$.
The effective field $\bm h_\mathrm{eff}(\bm r,t)$ is given by\cite{brown_micromagnetics_1963}:
\begin{equation}\label{eq:heff}
\bm h_\mathrm{eff}=-\frac{\delta g}{\delta \bm m}  \,\,,
\end{equation}
which takes into account interactions (exchange, anisotropy, magnetostatics, Zeeman) among magnetic moments and is expressed as the variational derivative of the free energy functional (the dimensionless energy is measured in units of $\mu_0 M_s^2 V$, with $V$ being the volume of region $\Omega$)
\begin{equation}\label{eq:free energy}
    g(\bm m,\bm h_a)=\frac{1}{V} \int_\Omega \frac{l_\mathrm{ex}^2}{2} (\nabla\bm m)^2 + f_\mathrm{an} - \frac{1}{2}\bm h_m\cdot \bm m - \bm h_a\cdot \bm m \,dV \,,
\end{equation}
where $A$ and $l_\mathrm{ex}=\sqrt{(2A)/(\mu_0 M_s^2)}$ are the exchange stiffness constant and length, respectively, $f_\mathrm{an}$ is the anisotropy energy density, $\bm h_m$ is the magnetostatic (demagnetizing) field and $\bm h_a(\bm r, t)$ the external applied field. 

When the anisotropy is of uniaxial type, such that $f_\mathrm{an}=\kappa_\mathrm{an} [1-(\bm m\cdot \bm e_\mathrm{an})^2]$ with $\kappa_\mathrm{an}$ and $\bm e_\mathrm{an}$ being the uniaxial anisotropy constant and unit-vector, respectively,  the effective field can be expressed by the sum of a linear operator $\mathcal{C}$ acting on magnetization vector field plus the applied field:
\begin{equation}
    \bm h_\mathrm{eff}(\bm r, t)= -\mathcal{C} \bm m + \bm h_a \,,
\end{equation}
where $\mathcal{C}=-l_\mathrm{ex}^2\nabla^2+\mathcal{N}+\kappa_\mathrm{an} e_\mathrm{an}\otimes e_\mathrm{an}$ and $\mathcal{N}$ is the (symmetric-positive definite) demagnetizing operator such that:
\begin{equation} \label{eq:demagnetizing field}
    \bm h_m(\bm r) = \frac{1}{4\pi}\nabla\nabla\cdot \int_\Omega \frac{\bm m(\bm r')}{|\bm r - \bm r'|} \, dV = -\mathcal{N}\bm m\,.
\end{equation}
As mentioned in the previous section, eq. \eqref{eq:iLLG} is usually complemented with the natural boundary conditions $\partial\bm m/\partial\bm n=0$ at the body surface $\partial\Omega$, which is typical when no surface anisotropy is considered.
It can be shown that the operator $\mathcal{C}$ with the aforementioned boundary conditions is self-adjoint and positive-definite in the appropriate subspace of square-integrable vector fields\cite{brown_micromagnetics_1963}.

It is also worth remarking that, for eq.\eqref{eq:iLLG}, equilibrium magnetization fields are characterized by simultaneously vanishing time-derivatives of first and second order:
\begin{equation}\label{eq:def equilibrium}
    \frac{\partial \bm m}{\partial t}=\bm 0 \quad,\quad \frac{\partial^2 \bm m}{\partial t^2}=\bm 0 \quad.
\end{equation}

Equation \eqref{eq:iLLG} describes a nonlinear dynamical system of higher order compared to that associated with the classical LLG equation (obtained by setting $\xi=0$ in eq.\eqref{eq:iLLG}). In fact, by defining a new variable $\bm w$ resembling, in a purely formal fashion, the 'angular momentum' of a point-particle of unitary  mass, position vector $\bm m$ and velocity $\partial\bm m /\partial t$ such that
\begin{equation}
\bm w=\bm m\times \frac{\partial \bm m}{\partial t} \,, \label{eq:definition w}   
\end{equation}
one has:
\begin{equation}
    \frac{\partial\bm w}{\partial t} = \bm m\times \frac{\partial^2\bm m}{\partial t^2} \quad.
\end{equation}
First, by dot-multiplying both sides of eq.\eqref{eq:iLLG} by $\bm m$, we observe that magnetization vector evolves on the unit-sphere $|\bm m|^2=1$ since 
\begin{equation}
\bm m\cdot \frac{\partial\bm m}{\partial t}=0 \,.   
\end{equation}
Then, by cross-multiplying both sides of eq.\eqref{eq:iLLG} by $\bm m$, one obtains:
\begin{equation}
    \bm w = -\bm m\times(\bm m \times \bm h_\mathrm{eff}) + \alpha \bm m\times \bm w + \xi \bm m\times\frac{\partial \bm w}{\partial t} \quad. 
\end{equation}
By performing further cross-multiplication of both sides of the latter equation by $\bm m$, one ends up with:
\begin{equation}\label{eq:pre iLLG w}
    \bm m\times\bm w = (\bm m \times \bm h_\mathrm{eff}) - \alpha  \bm w - \xi \frac{d\bm w}{dt} \quad, 
\end{equation}
where the property $\bm m\cdot \partial \bm w/ \partial t=0$ has been used.

Consequently, iLLG eq.\eqref{eq:iLLG} can be rewritten as a set two coupled nonlinear equations for variables $\bm m$ and $\bm w$ as follows:
\begin{align}
    \frac{\partial \bm m}{\partial t}&=\bm w\times \bm m \quad, \label{eq:iLLG_m}\\
    \xi\frac{\partial  \bm w}{\partial t}&=-\bm m\times \bm w -\alpha \bm w +\bm m\times \bm h_\mathrm{eff} \quad, \label{eq:iLLG_w}
\end{align}
where eq.\eqref{eq:iLLG_m} comes from eq.\eqref{eq:definition w} cross-multiplied by $\bm m$ combined with the fact that $|\bm m|^2=1$, and eq.\eqref{eq:iLLG_w} from eq.\eqref{eq:pre iLLG w}. 
We point out that, as a consequence of eq.\eqref{eq:def equilibrium} and the definition of $\bm w$ from eq.\eqref{eq:definition w}, equilibrium solutions of eqs.\eqref{eq:iLLG_m}-\eqref{eq:iLLG_w} are such that:
\begin{equation}\label{eq:def equilibrium w m}
   \frac{\partial \bm m}{\partial t}=\bm 0 \quad,\quad \frac{\partial \bm w}{\partial t}=\bm 0 \quad. 
\end{equation}
In this way, the implicit equation \eqref{eq:iLLG} has been transformed into a higher-order equation in standard explicit form, which is amenable of general considerations concerning the properties of the dynamical systems that it describes. 
 
To this end, we now focus on the dynamical system expressed by eqs.\eqref{eq:iLLG_m}-\eqref{eq:iLLG_w} where the state variables $\bm m,\bm w$ are considered independent of each other, remembering that it is equivalent to the original iLLG eq.\eqref{eq:iLLG} when eq.\eqref{eq:definition w} holds.
First of all, by dot-multiplying eq.\eqref{eq:iLLG_m} by $\bm m$, one can immediately see that the motion of vector $\bm m$ occurs on the unit-sphere $|\bm m|=1$:
 \begin{equation}\label{eq:amplitude preservation}
     \bm m\cdot \frac{\partial \bm m}{\partial t}=0 \quad \Rightarrow\quad |\bm m(\bm r,t)|=1 \quad\forall \bm r\in\Omega\,,\,t\geq t_0,
 \end{equation}
 provided that $\bm m$ has unit-amplitude at initial time $t_0$. The latter will be referred to as magnetization amplitude conservation property.
 
 Now, let us sum  eq.\eqref{eq:iLLG_m} dot-multiplied by $\bm w$ and eq.\eqref{eq:iLLG_w} divided by $\xi$ and dot-multiplied by $\bm m$. One has:
 \begin{equation}\label{eq:angular momentum projection conservation}
  \bm w\cdot \frac{\partial \bm m}{\partial t} + \bm m\cdot \frac{\partial \bm w}{\partial t}=\frac{\partial (\bm w\cdot \bm m)}{\partial t}= -\frac{\alpha}{\xi}\bm w\cdot\bm m \quad.   
 \end{equation}
 This means that, in any spatial location $\bm r\in\Omega$, the scalar product $\bm w\cdot\bm m$, termed as 'angular momentum' projection on magnetization, will have to decay exponentially to zero as follows:
\begin{equation}\label{eq:exponential wdotm}
    \bm w(\bm r,t)\cdot\bm m(\bm r,t)= \bm w(\bm r,t_0)\cdot\bm m(\bm r,t_0) e^{-\frac{\alpha}{\xi}t} \quad \forall \bm r\in\Omega\,,\,t\geq t_0,
\end{equation}
 where the time decay constant is controlled by the ratio $\xi/\alpha>0$ between the intensities of damping and inertia. Thus, for $t\gg t_0$ (practically $t>t_0+5\xi/\alpha$), the 'angular momentum' variable $\bm w$ is asymptotically constrained to evolve on the manifold defined by $\bm w\cdot \bm m=0$. Interestingly, for zero damping $\alpha=0$, the latter equation implies exact conservation of the product $\bm w\cdot\bm m$ at any time:
 \begin{equation}\label{eq:conservation wdotm}
    \bm w(\bm r,t)\cdot\bm m(\bm r,t)= \bm w(\bm r,t_0)\cdot\bm m(\bm r,t_0) \quad \forall \bm r\in\Omega\,,\,t\geq t_0.
\end{equation}
 
 From equation \eqref{eq:exponential wdotm} it is also worth noting that, for any value of $\alpha\geq 0$ and initially vanishing magnetization time-derivative $\partial \bm m /\partial t(\bm r,t_0)=0$ at any location $\bm r\in \Omega$, which therefore implies $\bm w(\bm r,t_0)=0$, the iLLG dynamics will occur such that the product $\bm w\cdot \bm m$ is always zero:
 \begin{equation}\label{eq:conservation wdotm zero}
    \bm w(\bm r,t)\cdot\bm m(\bm r,t)= \bm w(\bm r,t_0)\cdot\bm m(\bm r,t_0)=0 \quad \forall \bm r\in\Omega\,,\,t\geq t_0.
\end{equation}
 From the above discussion, being that the inertial magnetization dynamics must fulfill the two constraints \eqref{eq:amplitude preservation},\eqref{eq:exponential wdotm}, one can conclude that, in general, the dynamical system obtained by the iLLG eq.\eqref{eq:iLLG} and expressed by eqs.\eqref{eq:iLLG_m}-\eqref{eq:iLLG_w}  has, in each spatial location $\bm r\in\Omega$, four independent state variables evolving on a four-dimensional state space.  This means that the iLLG dynamics requires a double number of degrees of freedom compared to the classical LLG for its description.
 
 Furthermore, eq.\eqref{eq:iLLG} admits an additional conservation property. In fact, by dot-multiplying eq.\eqref{eq:iLLG_w} by $\bm w$ and integrating over the region $\Omega$, one has:
 \begin{align}
    \frac{1}{V}\int_\Omega \frac{\xi}{2}\frac{\partial |\bm w|^2}{\partial t} \,dV&= \frac{1}{V}\int_\Omega -\alpha |\bm w|^2 + \bm w\cdot(\bm m\times \bm h_\mathrm{eff}) \, dV \Leftrightarrow  \\
  \frac{1}{V}\int_\Omega \frac{\xi}{2}\frac{\partial |\bm w|^2}{\partial t} \, dV&= \frac{1}{V}\int_\Omega -\alpha |\bm w|^2 + \bm h_\mathrm{eff}\cdot\frac{\partial\bm m}{\partial t} \,dV. 
 \end{align}
By using the fact that 
\begin{equation}
    \frac{dg}{dt}=\frac{1}{V}\int_\Omega \frac{\delta g}{\delta\bm m}\cdot \frac{\partial\bm m}{\partial t} +\frac{\delta g}{\delta\bm h_a}\cdot \frac{\partial\bm h_a}{\partial t} \,dV = \frac{1}{V}\int_\Omega -\bm h_\mathrm{eff}\cdot \frac{\partial\bm m}{\partial t} -\bm m\cdot \frac{\partial\bm h_a}{\partial t} \,dV \,,
\end{equation}
and remembering from \eqref{eq:definition w} that $|\bm w|=|\partial\bm m/\partial t|$, one obtains the following energy balance equation:
\begin{align}\label{eq:energy balance}
    \frac{d}{d t}\left(g + \frac{1}{V}\int_\Omega\frac{\xi}{2}  \left|\bm w \right|^2 dV \right)&=-\frac{1}{V}\int_\Omega \bm m\cdot \frac{\partial \bm h_a}{\partial t} - \alpha\left|\bm w\right|^2 \, dV\Leftrightarrow  \nonumber\\
    \frac{d}{d t}\left(g + \frac{1}{V}\int_\Omega \frac{\xi}{2} \left|\frac{\partial \bm m}{\partial t} \right|^2 dV \right)&=-\frac{1}{V}\int_\Omega \bm m\cdot \frac{d\bm h_a}{d t} - \alpha\left|\frac{\partial \bm m}{\partial t}\right|^2 \,dV  \,\,.
\end{align}
The latter equation can be put in a more compact form by defining the following generalized free energy: 
\begin{equation}\label{eq:generalized energy}
    \tilde{g}(\bm m,\bm w,\bm h_a)=g(\bm m,\bm h_a) + \frac{1}{V}\int_\Omega \frac{\xi}{2} \left|\bm w \right|^2\, dV = g(\bm m,\bm h_a) + \frac{1}{V}\int_\Omega \frac{\xi}{2} \left|\frac{\partial\bm m}{\partial t} \right|^2\, dV  \quad,
\end{equation}
where the second term, in the framework of the purely formal mechanical analogy introduced before, can be seen as a sort of 'kinetic' energy (see the last equality in eq.\eqref{eq:generalized energy}) augmenting the classical micromagnetic free energy interpreted as 'potential' energy.
Thus, the balance equation \eqref{eq:energy balance} becomes
\begin{equation}\label{eq:generalized energy balance}
        \frac{d \tilde{g}}{d t}=-\frac{1}{V}\int_\Omega \bm m\cdot \frac{\partial\bm h_a}{\partial t} - \alpha\left|\frac{\partial \bm m}{\partial t}\right|^2 \,dV  \,\,,
\end{equation}
where the first term at the right-hand side describes energy pumping via time-varying external applied magnetic field and the second term takes into account the intrinsic dissipation of magnetic materials. 

It is apparent that, under the assumption of constant-in-time (even spatially-inhomogeneous) applied field  ($\partial\bm h_a/ \partial t=0$), the generalized free energy $\tilde{g}$ must be a decreasing function of time:
\begin{equation}\label{eq:Lyapunov structure}
        \frac{d \tilde{g}}{d t}= - \frac{1}{V}\int_\Omega\alpha\left|\frac{\partial \bm m}{\partial t}\right|^2 \,dV  \leq 0\,\,,
\end{equation}
which reveals a Lyapunov structure for the iLLG in terms of the generalized free energy $\tilde{g}$ similarly to what happens for the  LLG dynamics in terms of then classical free energy $g$. This means, that, under the above assumptions, the only possible attractors of the dynamics are stable equilibria  (i.e. such that $\partial\bm m /\partial t=0, \partial\bm w/\partial t=0$ and $\tilde{g}$ is minimum).

In addition, in the absence of dissipation ($\alpha=0$), one has the conservation property for the quantity $\tilde{g}$:
\begin{equation}\label{eq:conservation energy}
    \frac{d\tilde{g}}{dt}=\frac{d}{d t}\left(g + \frac{1}{V}\int_\Omega \frac{\xi}{2} \left|\frac{\partial\bm m}{\partial t} \right|^2\, dV \right) =0 \quad,
\end{equation}
which is analogous to the conservation of 'total' (potential + 'kinetic') energy $\tilde{g}$ in mechanical systems and here strikingly expresses the conservative nature of the (lossless) spin nutation dynamics.

We remark that the balance equation \eqref{eq:energy balance},\eqref{eq:generalized energy balance} could have been derived directly from eq.\eqref{eq:iLLG} by dot-multiplying both sides by the quantity in parentheses and integrating over $\Omega$. 

Finally, we observe that, in the absence of dissipation (i.e. $\alpha=0$), eqs.\eqref{eq:iLLG_m}-\eqref{eq:iLLG_w} admit three integrals of motion:
\begin{subequations}
    \begin{numcases}{}
        |\bm m(\bm r,t)| = 1 \quad \forall \bm r\in \Omega\,,\,t\geq t_0 \label{eq:iLLG amplitude conservation}  \\
        \bm w(\bm r,t)\cdot\bm m(\bm r,t) = \bm w(\bm r,t_0)\cdot\bm m(\bm r,t_0) \quad \forall \bm r\in \Omega\,,\,t\geq t_0 \label{eq:iLLG linear integral conservation} \\
        \tilde{g}=g + \frac{1}{V}\int_\Omega \frac{\xi}{2} \left|\frac{\partial\bm m}{\partial t} \right|^2\, dV = \tilde{g}_0 \,,\,t\geq t_0 \label{eq:iLLG energy conservation} 
   \end{numcases}
\end{subequations}
that we term amplitude, 'angular momentum' projection on magnetization and 'total' free energy conservation, respectively. The former two hold in a pointwise fashion, that is in any location and time instant (provided that they are fulfilled at initial time $t_0$), while the last is an integral constraint on magnetization motion (we remark that $\tilde{g}(t_0)=\tilde{g}_0$ is the initial 'total' free energy).        
   
The above conservation laws hold for spatially-inhomogeneous magnetization processes, but one can also consider 'sufficiently small' particles where the exchange interaction strongly penalizes spatial magnetization gradients and, thus, approximately treat them as
uniformly-magnetized (macrospin) anisotropic particles, which eliminates the dependence on the spatial location $\bm r$ within the ferromagnet. This makes sense  when dealing with magnetic nanosystems of dimensions in the order of the exchange length, such as those used as  elementary cells for magnetic memories and other spintronic devices\cite{dieny2020opportunities}. 
Under the assumption of spatially-uniform magnetization and anisotropy of uniaxial type, the free energy \eqref{eq:free energy} has the simple expression\cite{BMS2009}:
\begin{equation}\label{eq:free energy macrospin}
    g(\bm m, \bm h_a)=\frac{1}{2} D_x m_x^2 + \frac{1}{2} D_y m_y^2 +\frac{1}{2} D_z m_z^2-\bm m\cdot \bm h_a \quad,
\end{equation}
where $D_x,D_y,D_z$ are effective demagnetizing factors taking into account shape and crystalline anisotropy. 
The aforementioned integrals of motion \eqref{eq:iLLG amplitude conservation}-\eqref{eq:iLLG energy conservation} become
\begin{subequations}
    \begin{numcases}{}
        |\bm m(t)| = 1 \quad \forall \,,\,t\geq t_0 \,,\label{eq:iLLG amplitude conservation2}  \\
        \bm w(t)\cdot\bm m(t) =\bm w(t_0)\cdot\bm m(t_0) \quad \forall \,,\,t\geq t_0 \,,\label{eq:iLLG linear integral conservation2} \\
        \tilde{g}= g + \frac{\xi}{2} \left|\bm w \right|^2 =g + \frac{\xi}{2} \left|\frac{d\bm m}{d t} \right|^2 = \tilde{g}_0 \,,\,t\geq t_0 \,,\label{eq:iLLG energy conservation2} 
   \end{numcases}
   \end{subequations}
with $g$ given by eq.\eqref{eq:free energy macrospin} and will be instrumental in the validation of time-stepping techniques that we will discuss in the following sections.

\section{Spatially semi-discretized iLLG equation}

Now we proceed to the numerical discretization of the iLLG equation. 
 In the following, we will refer to spatially semi-discretized equations on a collection of $N$ mesh points $(\bm r_j)_{j=1}^N$ associated with the related computational cells of volume $V_j$. This description is quite general and works both for finite-difference and finite-element methods. 
 
 We will denote as $\underline{\bm m}(t)=(\bm m_1,\ldots , \bm m_N)^T$, $\underline{\bm w}(t)=(\bm w_1,\ldots, \bm w_N)^T$$ \in \mathbb{R}^{3N}$ (the notation $^T$ means matrix transpose) the mesh vectors containing all cell vectors $\bm m_j(t),\bm w_j(t) \in \mathbb{R}^3$ with $j=1,\ldots,N$.

Moreover, we will use the operator notation for the cross-product for both cell and mesh vectors, namely:
\begin{equation}
    \Lambda({\bm v})\cdot \bm w=\bm v\times \bm w\,,\quad \underline{\Lambda}({\underline{\bm v}})\cdot \underline{\bm w}=(\bm v_1\times \bm w_1, \ldots, \bm v_N\times \bm w_N)^T \,,
\end{equation}
meaning that the latter operator is a skew-symmetric $3N\times 3N$ block-diagonal operator that provides cross product of homologous cell vectors.

Thus, the semi-discretized iLLG equation will read as:
\begin{align}
    \frac{d \underline{\bm m}}{d t}&=\underline{\Lambda}(\underline{\bm w})\cdot \underline{\bm m} \quad, \label{eq:semidiscrete iLLG_m} \\
    \xi\frac{d  \underline{\bm w}}{d t}&=-\underline{\Lambda}(\underline{\bm m})\cdot \underline{\bm w} -\alpha \underline{\bm w} +\underline{\Lambda}({\underline{\bm m}})\cdot \underline{\bm h}_\mathrm{eff} \quad, \label{eq:semidiscrete iLLG_w}
\end{align}
where the discrete effective field $\underline{\bm h}_\mathrm{eff}$ is given by:
\begin{equation}\label{eq:discrete effective field}
    \underline{\bm h}_\mathrm{eff}(\underline{\bm m},t)=-\frac{\partial \underline{g}}{\partial\underline{\bm m}}=-C\cdot \underline{\bm m}(t) + \underline{\bm h}_a(t)\,,
\end{equation}
the symmetric positive-definite matrix $\underline{C}$ plays the role of the effective field operator $\mathcal{C}$ and $\underline{g}(\underline{\bm m},\underline{\bm h}_a)$ is the discrete counterpart of the free energy defined by eq. \eqref{eq:free energy}:
\begin{equation}\label{eq:discrete free energy}
    \underline{g}(\underline{\bm m},\underline{\bm h}_a)= \frac{1}{2} \underline{\bm m}^T\cdot \underline{C} \cdot \underline{\bm m} -\underline{\bm h}_a^T \cdot \underline{\bm m}\,.
\end{equation}
By using the same line of reasoning as for the continuous iLLG equation \eqref{eq:iLLG_m}-\eqref{eq:iLLG_w}, one can derive the following conservation properties:
\begin{align}
  &|\bm m_j(t)| = |\bm m_j(t_0)| \quad \forall t\geq t_0, \quad  j=1,\ldots,N \,, \label{eq:semidiscrete amplitude conservation} \\
&({\bm w_j(t)}\cdot{\bm m_j(t)})=({\bm w_j(t_0)}\cdot{\bm m_j(t_0)})e^{-\frac{\alpha}{\xi} t} \quad \forall t\geq t_0, \quad  j=1,\ldots,N \,, 
\label{eq:semidiscrete exponential wdotm} \\
  &\frac{d}{dt}\left( \underline{g}(\underline{\bm m}(t), \underline{\bm h}_a(t)) + \frac{\xi}{2}\left|\frac{d\underline{\bm m}}{dt}\right|^2 \right)= \frac{d}{dt}\left( \underline{g}(\underline{\bm m}(t), \underline{\bm h}_a(t)) + \frac{\xi}{2}\left|\underline{\bm w} \right|^2 \right)=\frac{d \underline{\tilde{g}}}{dt} = -\alpha \left|\frac{d\underline{\bm m}}{dt}\right|^2 - \underline{\bm m}\cdot \frac{d\underline{\bm h}_a}{dt} \,, \label{eq:semidiscrete energy balance} 
\end{align}
where $\underline{\tilde{g}}(\underline{\bm m},\underline{\bm w}, \underline{\bm h_a})=\underline{{g}}(\underline{\bm m},\underline{\bm h_a})+\frac{\xi}{2}|\underline{w}|^2$ is the discrete 'total' energy corresponding to $\tilde{g}$ in the continuous iLLG dynamics (see eq.\eqref{eq:generalized energy}).

\section{Midpoint time-steppings for iLLG dynamics}

The numerical solution of eqs.\eqref{eq:semidiscrete iLLG_m}-\eqref{eq:semidiscrete iLLG_w} with classical time-stepping techniques in general will corrupt the conservation properties \eqref{eq:semidiscrete amplitude conservation}-\eqref{eq:semidiscrete energy balance} of semi-discretized inertial magnetization dynamics. Thus, such properties will be fulfilled with an accuracy depending on the amplitude of the time-step $\Delta t$. For the classical purely precessional LLG equation, it has been shown\cite{dAquino2005geometrical}  that the implicit midpoint rule technique preserves the properties of discrete magnetization dynamics regardless of the time-step. Here we propose two schemes based on such technique for the iLLG spin nutation dynamics.   

\subsection{Implicit midpoint rule (IMR)}

The first is based on discretiztion of eqs.\eqref{eq:semidiscrete iLLG_m}-\eqref{eq:semidiscrete iLLG_w} at time $t^{n+\frac{1}{2}}=t^n+\Delta t/2$ with the following second-order midpoint formulas:
\begin{equation}\label{eq:mp formulas}
    \underline{\bm m}^{n+\frac{1}{2}}= \frac{\underline{\bm m}^{n+1}+ \underline{\bm m}^{n}}{2} \,\quad \underline{\bm w}^{n+\frac{1}{2}}= \frac{\underline{\bm w}^{n+1}+ \underline{\bm w}^{n}}{2} \,,
\end{equation}
where $\underline{\bm m}^{n}, \underline{\bm w}^{n}$ denote $\underline{\bm m}(t^{n}), \underline{\bm w}(t^{n})$, which leads to the following time-stepping for the $j-$th computational cell:
\begin{align}
    \frac{{\bm m_j}^{n+1}- {\bm m_j}^n}{\Delta t}&=\bm w_j^{n+\frac{1}{2}} \times \bm m_j^{n+\frac{1}{2}} \quad, \label{eq:iLLG_m_mp}\\
    \xi\frac{\bm w_j^{n+1}-\bm w_j^n}{\Delta t}&=-\bm m_j^{n+\frac{1}{2}}\times \bm w_j^{n+\frac{1}{2}} -\alpha \bm w_j^{n+\frac{1}{2}} \nonumber \\  &+ \bm m_j^{n+\frac{1}{2}}\times \bm h_\mathrm{eff,j}(\underline{\bm m}^{n+\frac{1}{2}},t^{n+\frac{1}{2}}) \quad,\quad j=1,\ldots,N\,. \label{eq:iLLG_w_mp}
\end{align}
Now, by dot-multiplying the first equation by $\bm m_j^{n+1/2}$, one can easily see that 
\begin{equation}
|\bm m_j^{n+1}|^2-|\bm m_j^{n}|^2=0 \,,\quad j=1,\ldots,N\,, \label{eq:IMP_ampl_cons}
\end{equation}
meaning that magnetization amplitude will be preserved unconditionally, namely independently of $\Delta t$ in each computational cell. 
In addition, by dot-multiplying eq.\eqref{eq:iLLG_m_mp} by $\bm w^{n+\frac{1}{2}}$ and eq.\eqref{eq:iLLG_w_mp} by $\bm m^{n+\frac{1}{2}}$ and summing both sides of the result, one immediately ends up with:
\begin{equation}\label{eq:IMP_exponential_wdotm}
 \frac{\bm w_j^{n+1}+\bm w_j^n}{2}\cdot \frac{\bm m_j^{n+1}-\bm m_j^n}{\Delta t} + \frac{\bm m_j^{n+1}+\bm m_j^n}{2}\cdot \frac{\bm w_j^{n+1}-\bm w_j^n}{\Delta t}  = -\frac{\alpha}{\xi} \bm w_j^{n+\frac{1}{2}}\cdot \bm m_j^{n+\frac{1}{2}} \,,\quad j=1,\ldots,N  \,,
\end{equation}
which expresses the reproduction of the property \eqref{eq:semidiscrete exponential wdotm} in its mid-point time discretized version:
\begin{equation}\label{eq:IMP_exponential_wdotm2}
 \frac{\bm w_j^{n+1}\cdot\bm m_j^{n+1}-\bm w_j^n\cdot \bm m_j^n}{\Delta t}= -\frac{\alpha}{\xi} \bm w_j^{n+\frac{1}{2}}\cdot \bm m_j^{n+\frac{1}{2}} \,,\quad j=1,\ldots,N   \,.
\end{equation}
Remarkably enough, in the conservative case $\alpha=0$, the latter equation becomes:
\begin{equation}\label{eq:IMP_exponential_wdotm3}
 \frac{\bm w_j^{n+1}\cdot\bm m_j^{n+1}-\bm w_j^n\cdot \bm m_j^n}{\Delta t}= 0 \quad\Rightarrow\quad \bm w_j^{n+1}\cdot \bm m_j^{n+1}=\bm w_j^{n}\cdot \bm m_j^{n}\,,\quad j=1,\ldots,N   
\end{equation}
which means that the 'angular momentum' projection conservation property is fulfilled for any choice of the time step $\Delta t$. 

Now let us consider the midpoint rule time-stepping for the mesh vectors:
\begin{align}
    \frac{{\underline{\bm m}}^{n+1}- {\underline{\bm m}}^n}{\Delta t}&=\underline{\Lambda}(\underline{\bm w}^{n+\frac{1}{2}}) \cdot \underline{\bm m}^{n+\frac{1}{2}} \quad, \label{eq:mesh iLLG_m_mp}\\
    \xi\frac{\underline{\bm w}^{n+1}-\underline{\bm w}^n}{\Delta t}&=-\underline{\Lambda}(\underline{\bm m}^{n+\frac{1}{2}})\cdot \underline{\bm w}^{n+\frac{1}{2}} -\alpha \underline{\bm w}^{n+\frac{1}{2}} + \underline{\Lambda}(\underline{\bm m}^{n+\frac{1}{2}})\cdot \underline{\bm h}_\mathrm{eff}(\underline{\bm m}^{n+\frac{1}{2}},t^{n+\frac{1}{2}}) \quad. \label{eq:mesh iLLG_w_mp}
\end{align}

By assuming constant applied field, dot-multiplying the second equation by $\underline{\bm w}^{n+1/2}$ and taking into account eq.\eqref{eq:discrete effective field} and the symmetry of the matrix $\underline{C}$, one obtains the following discretized energy balance:
\begin{align}
    \frac{\xi}{2}\frac{|\underline{\bm w}^{n+1}|^2-|\underline{\bm w}^n|^2}{\Delta t}&= -\alpha \bm |\underline{\bm w}^{n+\frac{1}{2}}|^2 - \frac{g^{n+1}-g^n}{\Delta t} \Leftrightarrow  \\ \nonumber
    \frac{\underline{\tilde{g}}^{n+1}-\underline{\tilde{g}}^n}{\Delta t}&= -\alpha \bm |\underline{\bm w}^{n+\frac{1}{2}}|^2 \quad, \label{eq:IMP_en_bal} 
\end{align}
where $\underline{\tilde{g}}^n=\underline{\tilde{g}}(\underline{\bm m}^n)$, which implies that the total (discrete) energy $\underline{\tilde{g}}$ must be either decreasing when $\alpha>0$ or being conserved when $\alpha=0$, both regardless of the time-step.

Equations \eqref{eq:iLLG_m_mp}-\eqref{eq:iLLG_w_mp} represent a nonlinear system of $6N$ coupled scalar equations, which must be solved at each time step. They can be regarded as the following two vector equations in $\underline{\bm u}=\underline{\bm m}^{n+1}, \underline{\bm v}=\underline{\bm w}^{n+1}$:
\begin{equation}
    \bm F(\underline{\bm u},\underline{\bm v})=\bm 0 \quad,\quad \bm G(\underline{\bm u},\underline{\bm v}) =\bm 0 \,,
\end{equation}
and can be solved by using Newton-Raphson iteration:
\begin{align}
    \left( \begin{array}{c} \underline{\bm u}_{k+1}  \\ \underline{\bm v}_{k+1} \end{array}\right) = \left( \begin{array}{c} \underline{\bm u}_{k}  \\ \underline{\bm v}_{k} \end{array} \right) -
    \left( \begin{array}{cc} 
    \frac{\partial \bm F}{\partial \underline{\bm u}} & \frac{\partial \bm F}{\partial \underline{\bm v}} \\ 
    \frac{\partial \bm G}{\partial \underline{\bm u}} & \frac{\partial \bm G}{\partial \underline{\bm v}} \end{array} \right)^{-1} \cdot  \left( \begin{array}{c} \bm F(\underline{\bm u}_{k})  \\ \bm G (\underline{\bm v}_{k}) \end{array}\right) \,,
\end{align}
where the partial Jacobian matrices are given by:
\begin{align}
    \frac{\partial \bm F}{\partial \underline{\bm u}} &= \frac{\underline{I}}{\Delta t} -\frac{1}{4}\underline{\Lambda}(\underline{\bm v}+\underline{\bm w}^n) \,,\label{eq:IMR Jacobian1}\\
    \frac{\partial \bm F}{\partial \underline{\bm v}} &=  \frac{1}{4}\underline{\Lambda}(\underline{\bm u}+\underline{\bm m}^n) \,,\\
    \frac{\partial \bm G}{\partial \underline{\bm u}} &=  -\frac{1}{4}\underline{\Lambda}(\underline{\bm v}+\underline{\bm w}^n) + \frac{1}{2}\underline{\Lambda}(\underline{\bm u}+\underline{\bm m}^n)\cdot \underline{C} - 
    \underline{\Lambda}\left[\underline{\bm h}_\mathrm{eff}\left(\frac{\underline{\bm u}+\underline{\bm m}^n}{2}\right)\right] \,, \label{eq:IMR Jacobian3}\\
        \frac{\partial \bm G}{\partial \underline{\bm v}} &= \frac{\xi }{\Delta t}\,\underline{I} +\frac{1}{4}\underline{\Lambda}(\underline{\bm u}+\underline{\bm m}^n) + \frac{\alpha}{2}\,\underline{I} \,, \label{eq:IMR Jacobian4}
\end{align}
and the linear operator notation $\underline{\Lambda}$ has been used for the cross product involving mesh vectors.

The above time-stepping has remarkable qualitative properties that reproduce those of the continuous iLLG dynamics and, therefore, represents the preferred choice to realize inertial micromagnetic numerical codes for the analysis of terahertz magnetization dynamics.  

However, it evidently requires to double the state variables and, consequently, the number of unknowns, owing to the introduction of the vector field $\bm w$ although one is mainly interested to compute the dynamics of magnetization vector field $\bm m$. This issue becomes even more pronounced when large-scale micromagnetic simulations with full spatial discretization are considered, which would require dramatic modification of numerical codes in order to introduce the auxiliary variable $\bm w$ and to solve a system of $6N$ nonlinear coupled equations at each time-step. Moreover, we remark that in the latter situation, the $3N\times 3N$ matrix $\underline{C}$ involved in the Newton iteration (see eq.\eqref{eq:IMR Jacobian3}) is also fully-populated owing to the long-range nature of magnetostatic interactions. Therefore, following what is done for the classical LLG equation\cite{dAquino2005geometrical}, a quasi-Newton technique is required to solve the large nonlinear system, implemented by considering reasonable and sparse approximation of the matrix $\underline{C}$ (e.g. obtained retaining only exchange and anisotropy contributions $\underline{C}_\mathrm{ex}$ and $\underline{C}_\mathrm{an}$, respectively) and, in turn, of the full Jacobian defined by eqs.\eqref{eq:IMR Jacobian1}-\eqref{eq:IMR Jacobian4}. Of course, the computational cost of such quasi-Newton method, involving the solution of several non-symmetric large linear systems (e.g. by using GMRES methods\cite{Saad1986}), will be at least double with respect to that associated with LLG equation, posing a strong limit to the capability of solving large-scale iLLG dynamics.

\subsection{Implicit midpoint with multi-step inertial term (IMR-MS)}

For these reasons, in order to obtain an alternative efficient numerical technique with minimum modification of existing micromagnetic codes, we propose a second time-stepping based on the implicit midpoint rule combined with a multi-step method for the inertial term. This technique is based on direct space-time discretization of eq.\eqref{eq:iLLG} at time $t^{n+\frac{1}{2}}=t_n+\Delta t/2$:
\begin{align}
    \frac{\underline{\bm m}^{n+1}- \underline{\bm m}^n}{\Delta t}&= - \underline{\Lambda}(\underline{\bm m}^{n+\frac{1}{2}})\cdot \left(\underline{\bm h}_\mathrm{eff}\left(\underline{\bm m}^{n+\frac{1}{2}},t^{n+\frac{1}{2}}\right) \right. -\alpha \frac{\underline{\bm m}^{n+1}- \underline{\bm m}^n}{\Delta t}
    -\xi \frac{d^2\underline{\bm m}}{dt^2}\left|_{n+\frac{1}{2}} \right) \quad. \label{eq:iLLG_mp}
\end{align}

Then, in order to retain the amplitude conservation property \eqref{eq:IMP_ampl_cons} of the aforementioned IMR scheme, we use the first of midpoint formulas \eqref{eq:mp formulas} in eq.\eqref{eq:iLLG_mp} in a way that it is rewritten as system of $3N$ nonlinear equations in the unknowns $\underline{\bm m}^{n+1}$:
\begin{equation}\label{eq:nonlinear_system}
\bm F^n(\underline{\bm m}^{n+1})=\mathbf{0}\quad,
\end{equation}
where $\bm F^n(\underline{\bm y}):\mathbb{R}^{3N}\rightarrow \mathbb{R}^{3N}$
is the following vector function:
\begin{equation}\label{eq:nonlinear_function}
\bm F^n(\underline{\bm y})=\left[I-\alpha\underline{\Lambda}\left(\frac{\underline{\bm y}+\underline{\bm m}^{n}}{2}\right)\right]
\left(\underline{\bm y}-\underline{\bm m}^{n}\right) - \Delta t
\,\,\bm f^n\left(\frac{\underline{\bm y}+\underline{\bm m}^{n}}{2}\right) -\Delta t\, \xi \frac{d^2\underline{\bm m}}{dt^2}\left|_{n+\frac{1}{2}} \right) \,,
\end{equation}
and where
\begin{equation}\label{eq:f_conservative_flow}
   \bm f^n(\underline {\bm m})=-\underline{\Lambda}(\underline{\bm m}) \cdot \underline{\bm h}_\text{eff}\left(\underline{\bm m},t^n+\frac{\Delta t}{2}\right)=  \underline{\Lambda}(\underline{\bm m}) \cdot
   \frac{\partial \underline{g}}{\partial \underline{\bm m}}\left(\underline{\bm m},\underline{\bm h}_a \left(t^n+\frac{\Delta t}{2}\right)\right) 
\end{equation}
is the purely precessional term in the right-hand-side of the conservative semi-discretized iLLG equation \eqref{eq:semidiscrete iLLG_w} expressed by using the definition \eqref{eq:discrete effective field} of the discrete effective field.

Then, we adopt a multi-step approach with a $p-$points backward formula for the second derivative in the inertial term appearing in eqs.\eqref{eq:iLLG_mp} and \eqref{eq:nonlinear_function}:
\begin{equation}\label{eq:d2m multi step}
    \frac{d^2\underline{\bm m}}{dt^2}\left|_{n+\frac{1}{2}} \right. \approx
\frac{1}{\Delta t^2}\sum_{k=1}^{p\geq 3} a_{n+2-k} \underline{\bm m}^{n+2-k} =\Delta^2_{p}\,,
\end{equation}
where the coefficients $a_{n+2-k}$ are determined from truncation error analysis in order to control the accuracy of the approximation.
This technique implies a slight modification of existing numerical codes based on implicit midpoint rule time-stepping. In fact, once formula \eqref{eq:d2m multi step} is plugged into the time-stepping equation \eqref{eq:iLLG_mp}, the solution of the nonlinear coupled equations \eqref{eq:nonlinear_system} can be obtained by using the Newton-Raphson technique\cite{dAquino2005geometrical} as follows:
\begin{equation}
\underline{\bm y}_0=\underline{\bm m}^n \label{eq:iter} \,  , \,\,\,\,
\underline{\bm y}_{k+1}=\underline{\bm y}_k+\Delta\underline{\bm y}_{k+1} \quad  \text{with} \quad
J^n_F(\underline{\bm y}_k,t^n)\Delta\underline{\bm y}_{k+1}=-\bm F^n(\underline{\bm y}_k) \,,
\end{equation}
by simply considering the following augmented Jacobian matrix of the iteration:
\begin{align}
    J_F(\underline{\bm u},t)&=\frac{\underline{I}}{\Delta t}+\frac{\alpha}{\Delta t}\underline{\Lambda}(\underline{\bm m}^n)+\frac{\xi}{2\Delta t^2}a_1\underline{\Lambda}(\underline{\bm u}) +\frac{\xi}{2\Delta t^2}\underline{\Lambda}\left(\sum_{k=2}^{p\geq 3} a_{n+2-k} \underline{\bm m}^{n+2-k}\right)- \nonumber \\&-\frac{\xi}{2\Delta t^2}\underline{\Lambda}(\underline{\bm u} + \underline{\bm m}^n) a_1-\frac{1}{2}J_f\left(\frac{\underline{\bm u}+ \underline{\bm m}^n}{2},t+\frac{\Delta t}{2}\right)
\end{align}
where $J_f(\underline{\bm u},t)=\underline{\Lambda}(\underline{\bm u})\cdot \underline{C}+\underline{\Lambda}[\underline{\bm h}_\mathrm{eff}(\underline{\bm u},t)]$.
The linear system in eq.\eqref{eq:iter} is solved at each iteration $k$ by considering the sparse approximation of the full matrix $\underline{C}$ as $\underline{C}\approx \underline{C}_\mathrm{ex}+\underline{C}_\mathrm{an}$ in the Jacobian $J_f$ and using the GMRES method\cite{Saad1986} until the residual $\|\bm F^n(\underline{\bm y}_k)\|$ decreases below a prescribed tolerance. 

\begin{table}[t]
    \centering
    \begin{tabular}{|c|c|c|c|}
    \hline 
    $p$ &order & Coefficients & scheme \\
    \hline
    3 & $\mathcal{O}(\Delta t)$    & $a_1=1, a_0=-2, a_{-1}=1$ & IMR-MS1\\
    4 & $\mathcal{O}(\Delta t^2)$    & $a_1=3/2, a_0=-7/2, a_{-1}=5/2, a_{-2}=-1/2$ & IMR-MS2\\
     \hline
    \end{tabular}
    \caption{Table of coefficients for multi-step formula \eqref{eq:d2m multi step}.}
    \label{tab:tab1}
\end{table}

Now, if one assumes that initially magnetization has zero velocity $d\underline{\bm m}/dt(t=0)=\bm 0$, which is reasonable in simulation of experimental situations, then at the first time step $n=1$ one has $\bm m^{n+2-k}=0, k>2$. For the subsequent steps $n>1$, one can use magnetization samples from the previous steps as $\underline{\bm m}^{n+2-k}, k>2$ in eq.\eqref{eq:iLLG_mp}. The only cost of such operation is for the storage of $p-2$ magnetization vectors.

In this respect, the simplest choice is the classical $p=3$ points formula:
\begin{equation} \label{eq:D2p3}
    \frac{d^2\underline{\bm m}}{dt^2}\left|_{n+\frac{1}{2}} \right. \approx \frac{\underline{\bm m}^{n+1} + \underline{\bm m}^{n-1}- 2 \underline{\bm m}^n}{\Delta t^2} = \Delta^2_{p=3}\,,
\end{equation}
which, plugged into eq.\eqref{eq:iLLG_mp}, defines the IMR-MS1 scheme.

However, an analysis of truncation error reveals that:
\begin{equation}
      \Delta^2_{p=3}=\frac{d^2\underline{\bm m}}{dt^2}\left|_{n+\frac{1}{2}} \right. - \frac{1}{2} \frac{d^3\underline{\bm m}}{dt^3}\left|_{n+\frac{1}{2}}\right. \Delta t + \frac{5}{24} \frac{d^4\underline{\bm m}}{dt^4}\left|_{n+\frac{1}{2}}\right. \Delta t^2 + \ldots \,,
\end{equation}
meaning that the accuracy is just of first order $\mathcal{O}(\Delta t)$ (it would be $\mathcal{O}(\Delta t^2)$ if the second derivative was computed at $t=t_n$). Thus, in order to be consistent with discretization of other terms in eq.\eqref{eq:iLLG_mp} at second order with respect to $\Delta t$, one can derive a more accurate formula using $p=4$ points. To this end, we compute a different second derivative formula:
\begin{equation}
    \tilde{\Delta}^2_{p=3}=\frac{d^2\underline{\bm m}}{dt^2}\left|_{n+\frac{1}{2}} \right. \approx \frac{2\underline{\bm m}^{n+1}-3\underline{\bm m}^n+\underline{\bm m}^{n-2}}{3\Delta t^2} \,,
\end{equation}
which has truncation error such that:
\begin{equation}
         \tilde{\Delta}^2_{p=3}=\frac{d^2\underline{\bm m}}{dt^2}\left|_{n+\frac{1}{2}} \right. - \frac{5}{6} \frac{d^3\underline{\bm m}}{dt^3}\left|_{n+\frac{1}{2}}\right. \Delta t + \frac{13}{24} \frac{d^4\underline{\bm m}}{dt^4}\left|_{n+\frac{1}{2}}\right. \Delta t^2 + \ldots \,. 
\end{equation}
Now we use Richardson extrapolation\cite{richardson1911} to cancel $\mathcal{O}(\Delta t)$ order terms in the truncation and define the following new difference formula (defining the IMR-MS2 scheme):
\begin{equation} \label{eq:D2p4}
    \Delta^2_{p=4}=\frac{5}{2}\Delta^2_{p=3} - \frac{3}{2}\tilde{\Delta}^2_{p=3} = \frac{3\underline{\bm m}^{n+1}-7\underline{\bm m}^n+5 \underline{\bm m}^{n-1}-\underline{\bm m}^{n-2}}{2\Delta t^2} \,,
\end{equation}
for which the truncation error is $\mathcal{O}(\Delta t^2)$:
\begin{equation}
           {\Delta}^2_{p=4}=\frac{d^2\underline{\bm m}}{dt^2}\left|_{n+\frac{1}{2}} \right. - \frac{7}{24} \frac{d^4\underline{\bm m}}{dt^4}\left|_{n+\frac{1}{2}}\right. \Delta t^2 + \ldots \,.  
\end{equation}
The coefficients for the above multi-step formulas with $p=3,4$ are summarized in table \ref{tab:tab1}.

In order to assess the order of accuracy of the proposed schemes, we consider the conservative iLLG dynamics ($\alpha=0$) and numerically integrate eq.\eqref{eq:iLLG} using IMR, IMR-MS1, IMR-MS2 time-steppings for different time step $\Delta t$ and compare the results with a benchmark reference solution obtained by using standard adaptive time step Dormand-Prince Runge-Kutta (RK45) scheme\cite{dormand1980,shampine1997}. Absolute tolerances are set to $10^{-14}$ both for RK45 and for Newton iterations solving eq.\eqref{eq:iLLG_mp} with IMR, IMR-MS1, IMR-MS2. 

\begin{figure}[t]
    \centering
 \includegraphics[width=0.49\textwidth]{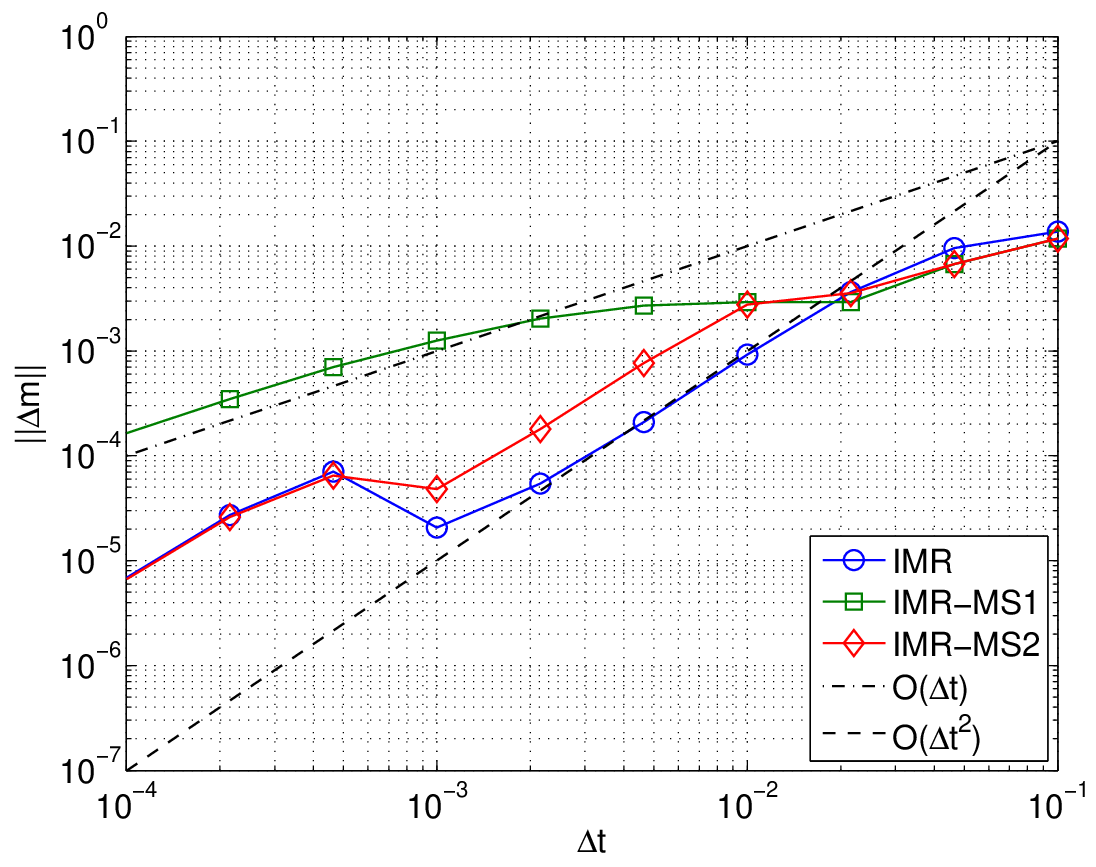}   \includegraphics[width=0.49\textwidth]{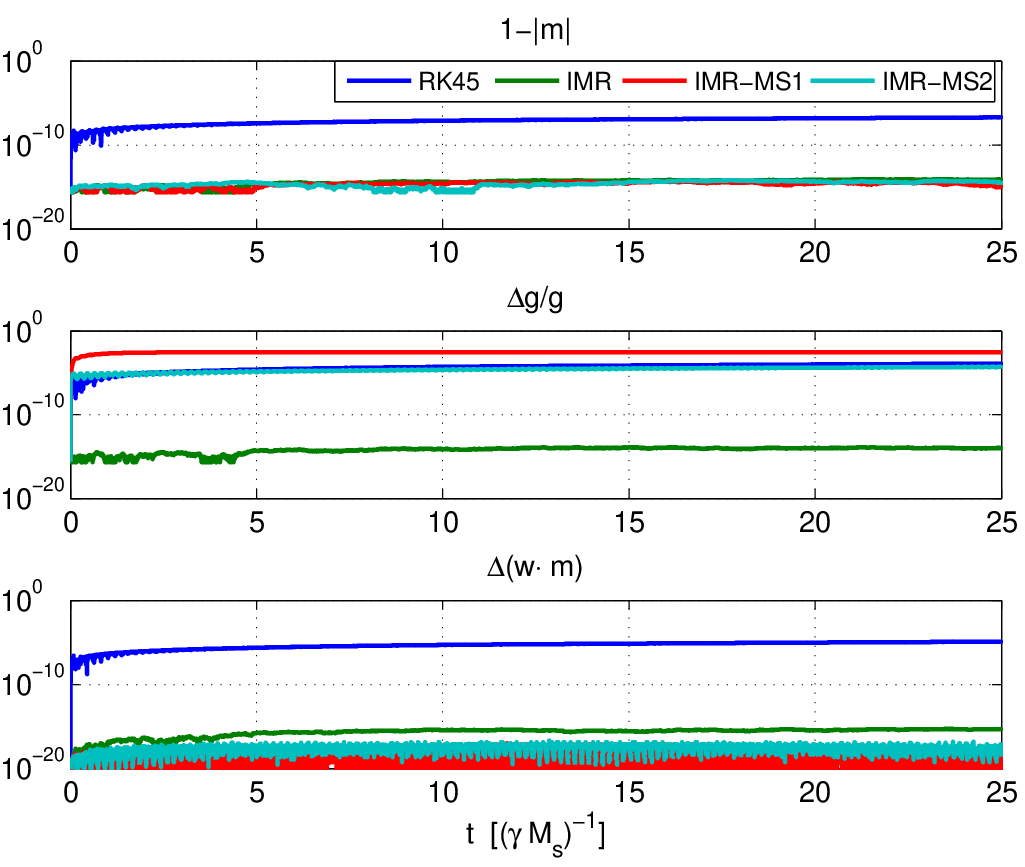} 
   
    \caption{Accuracy tests on conservative ($\alpha=0$) iLLG dynamics. The values of parameters are $D_x=0,1,D_y=0.2,D_z=0.7,\bm h_a=(0,0,0.1), \bm m(t=0)=\bm (1,0,0), \xi=0.03$. (left) Global error $||\Delta\bm m||$  at $t=1$ between IMR, IMR-MS1, IMR-MS2 schemes and reference RK45 solution showing first-order $O(\Delta t)$ behavior for IMR-MS1 and second-order $O(\Delta t^2)$ for IMR and IMR-MS2. (right) Conservation of properties of iLLG dynamics versus time (time step $\Delta t=0.001$ for all IMR schemes). Top panel refers to amplitude $1-|\bm m|$ conservation, middle panel to relative error $\Delta\tilde{g}/\tilde{g}=(\tilde{g}(t)-\tilde{g}(0))/\tilde{g}(0)$ in 'total' free energy  conservation, bottom panel refers to error $\Delta(\bm w \cdot\bm m)=\bm w(t)\cdot\bm m(t)-\bm w(0)\cdot\bm m(0)$ in 'angular momentum' projection on magnetization  conservation. One can see that all IMR, IMR-MS1, IMR-MS2 schemes preserve amplitude $|\bm m|$ and projection $\bm w\cdot\bm m$ with (double-precision) machine accuracy and only IMR also preserves energy. IMR-MS2 outperforms IMR-MS1 and RK45 in energy conservation.}
    \label{fig:conservative iLLG test}
\end{figure}

In the left panel of figure \ref{fig:conservative iLLG test}, we report the global truncation error $||\Delta\bm m||$ arising from time-integration of iLLG eq.\eqref{eq:iLLG} in the interval [0,1] between the proposed schemes and the reference RK45 solution. A quick inspection of the figure confirms the expected first and second orders of accuracy for IMR-MS1 and IMR,IMR-MS2, respectively. Remarkably, IMR-MS2 has performance quite similar to the fully-implicit IMR without doubling the number of degrees of freedom. On the other hand, in the right panel of fig.\ref{fig:conservative iLLG test}, one can look at the conservation properties for the proposed schemes (with $\Delta t=0.001$) and the reference solution. As expected, all the three IMR schemes are able to preserve amplitude $|\bm m|$ and 'angular momentum' projection on magnetization $\bm w\cdot\bm m$ with (double) machine precision, while only the fully-implicit IMR is able to do so for the 'total' energy $\tilde{g}$. Nevertheless, it is apparent (middle panel, blue and cyan solid lines) that IMR-MS2 is able to guarantee the same precision as the RK45 concerning energy conservation while being significantly lower-order than RK45. For the evaluation of the variable $\bm w$ and energy $\tilde{g}$ when considering IMR-MS schemes, we have used the central difference formula $\bm w^n=\bm m^n \times (\bm m^{n+1}-\bm m^{n-1})/(2\Delta t)$. 

\section{Numerical results}

In order to validate the proposed techniques on physically relevant situations, we perform two different simulations. The first describes the ultra-fast resonant spin nutation of a uniformly-magnetized thin-film driven by ac terahertz appled field, similar to the experiment\cite{neeraj2021inertial} that provided direct evidence of the presence of inertial effects. This will also be a basic testbed to compare the accuracy of the developed IMR and IMR-MS schemes. The second simulation will address the ultra-fast spatially-inhomogeneous dynamics of magnetization in a microscale nanodot excited with terahertz applied field and will demonstrate the efficiency of the IMR-MS time-stepping in full micromagnetic simulations.

\subsection{Nutation frequency response of a single-domain particle}

We analyze the frequency response of a thin-film magnetized along the easy $y$ direction and subject to an out-of-plane ac field with small amplitude. In this situation, one can assume that macrospin iLLG dynamics occurs in the linear regime and analytical theory can be developed. To this end, let us assume that the applied field is decomposed in a nonzero constant bias field plus a time-harmonic component $\bm h_a(t)=\bm h_\mathrm{dc}+\bm h_\mathrm{ac}(t)\,\,, \quad |\bm h_\mathrm{ac}|\ll|\bm h_\mathrm{dc}|$. We also assume that the free energy has the simple form \eqref{eq:free energy macrospin} under the macrospin approximation.

Then, the iLLG eq.\eqref{eq:iLLG} can be linearized around an equilibrium $\bm m_0$ such that $\bm m(t)=\bm m_0 +\Delta\bm m(t)$ in the following way:
\begin{equation}\label{eq:linear iLLG}
    \frac{d \Delta \bm m}{d t}=\bm m_0 \times \left[(D+h_0\mathcal{I}) \cdot\Delta \bm m - \alpha\frac{d \Delta \bm m}{d t} - \xi \frac{d^2 \Delta \bm m}{d t^2}\right] \,\,,
\end{equation}
where $D$ denotes the diagonal matrix $D=\text{diag}[D_x,D_y,D_z]$, $h_0=\bm h_\mathrm{eff}(\bm m_0)\cdot \bm m_0 =(-D\cdot \bm m_0 +\bm h_a) \cdot \bm m_0$ is the projection of the equilibrium effective field on equilibrium magnetization.

We first observe that the dynamics fulfills the constraint $\bm m_0\cdot\Delta\bm m=0$ (to the first order), therefore we can consider only the dyanamics of the component $\Delta \bm m_\perp$ of $\Delta \bm m$ living in the plane perpendicular to the equilibrium $\bm m_0$. We also refer to the projection of the demag tensor $D$ as $D_\perp$. As a consequence, we will deal with vectors having only two components associated with axes transverse to the equilibrium $\bm m_0$. 
We observe that the skew-symmetric operator $\Lambda$ is invertible  (such that $\Lambda\cdot\Lambda=-\mathcal{I}$) when restricted to the plane orthogonal to $\bm m_0$ and we express both $\Delta \bm m_\perp,\bm h_\mathrm{ac}$ using complex (phasor) domain as $\Delta \bm m_\perp(t)=\Delta  \tilde{\bm m} e^{j\omega t} \,\,,\,\, \bm h_\mathrm{ac}=\tilde{\bm h}_\mathrm{ac} e^{j\omega t}$.
By using these formulas in eq.\eqref{eq:linear iLLG}, one ends up with:
\begin{equation}\label{eq:iLLG an freq resp}
    \Delta \tilde{\bm m}= \underbrace{\left[j\omega\Lambda(\bm m_0)+(D_\perp+h_0\mathcal{I})+j\omega\alpha\mathcal{I}-\xi\omega^2\mathcal{I}\right]^{-1}}_{\chi(\omega)} \tilde{\bm h}_\mathrm{ac} \,\,,
\end{equation}
which defines the magnetic susceptibility tensor $\chi(\omega)$.
When referred to principal axes, $\chi(\omega)$ is a $2\times 2$ matrix which can be easily computed.

\begin{figure}
    \centering
    \includegraphics[width=0.6\textwidth]{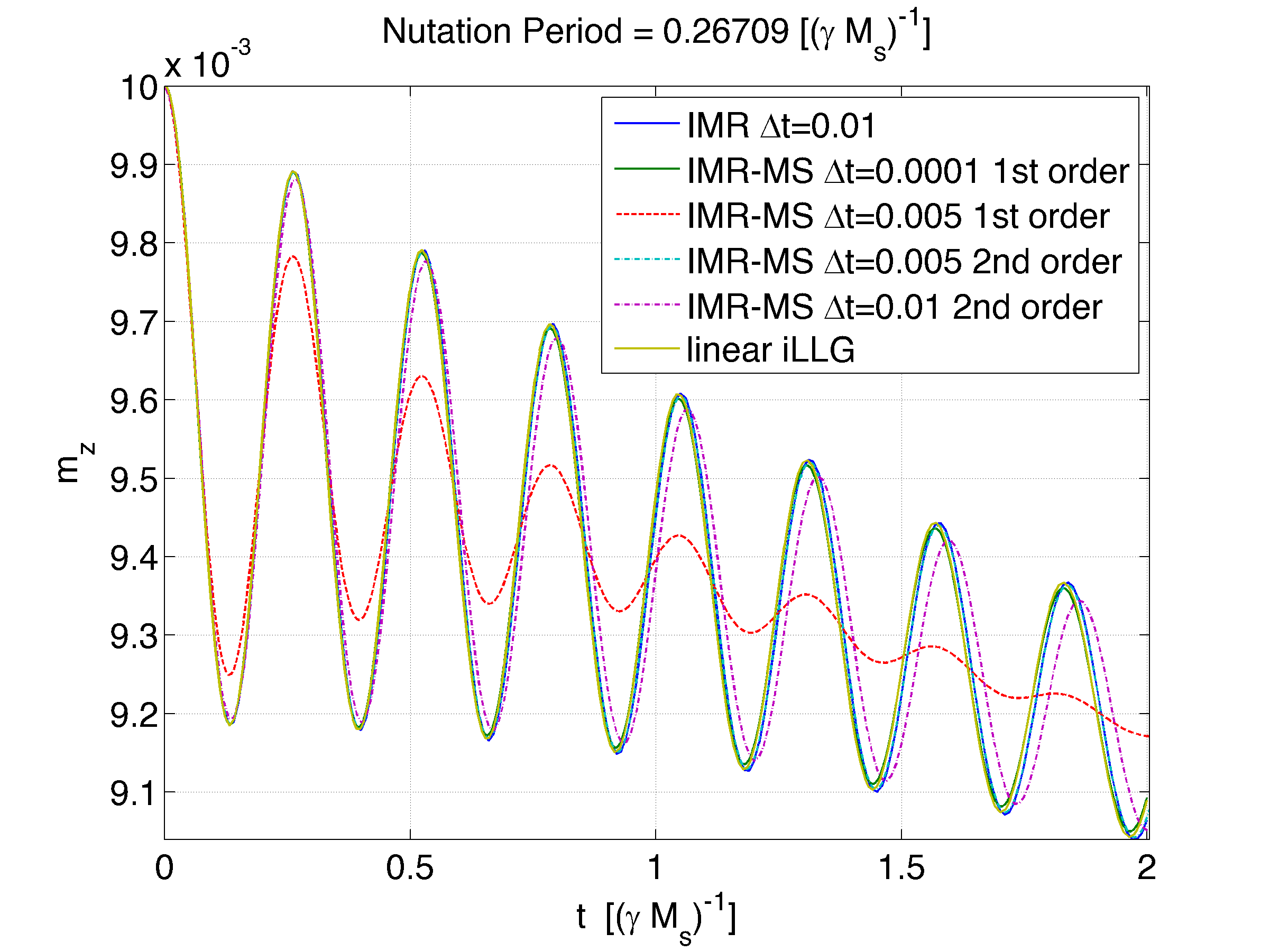}
   
    \caption{Time-domain linear relaxation of $m_z$ computed with IMR, IMR-MS with different time steps and compared with analytical solution of linear iLLG eq.\eqref{eq:linear iLLG}.}
    \label{fig:lin iLLG time domain}
\end{figure}

It can be shown that the resonance frequencies associated with the above linear dynamical system are the roots of the following fourth-degree polynomial:
\begin{equation}\label{eq:characteristic polynomial}
  \xi^2\omega^4-2j\alpha \xi \omega^3-(\alpha^2+\xi(\omega_{0y}+\omega_{0z}) +1)\omega^2  + j\alpha(\omega_{0y}+\omega_{0z})\omega+\omega_{0y}\omega_{0z}=0 \,\,,
\end{equation}
where $\omega_{0y}=D_y-D_x+h_\mathrm{dc}$ and $\omega_{0y}=D_z-D_x+h_\mathrm{dc}$.
Equation \eqref{eq:characteristic polynomial} can be solved by using appropriate perturbation theory leading to the following resonance frequencies (computed in the conservative case when $\alpha=0$):
\begin{align}\label{eq:resonance frequencies}
    \omega_\mathrm{FMR}&\approx \pm\frac{\omega_K}{\sqrt{1+\xi(\omega_{0y}+\omega_{0z}})} \,\,,\,\,\omega_K=\sqrt{\omega_{0y}\omega_{0z}} \,\,,\\
    \omega_N &\approx \pm \sqrt{ \frac{\sqrt{2\xi(\omega_{0y}+\omega_{0z})+1} + \xi(\omega_{0y}+\omega_{0z})+1}{2\xi^2} } \,\,, \label{eq:resonance frequencies nut}
\end{align}
where $\omega_K=\sqrt{(D_y-D_x+h_\mathrm{dc})(D_z-D_x+h_\mathrm{dc})}$ is the classical Kittel ferromagnetic resonance (FMR) frequency. 
The former equation describes the influence of inertial effects on the FMR frequency, while the second formula gives the nutation resonance frequency (typically in the THz range).  We observe that the above formulas take into account the dependence on the external bias field through the parameters $\omega_{0y},\omega_{0z}$.
It is also possible to determine closed-form expressions for the half-power (Full Width at Half Maximum, FWHM) linewidths:
\begin{align}\label{eq:resonance linewidths}
    \Delta \omega_\mathrm{FMR} &\approx \alpha(\omega_{0y}+\omega_{0z}) -\xi(\omega_{0y}^2+4\omega_{0y}\omega_{0z}+\omega_{0z}^2 )  \,\,, \\
    \Delta \omega_N &\approx \frac{\alpha}{\xi}\left[1+\frac{1}{\sqrt{2\xi(\omega_{0y}+\omega_{0z})+1}} \right] \,\,. \label{eq:resonance linewidths nut}
\end{align}

\begin{figure}
    \centering
    \includegraphics[width=0.5\textwidth]{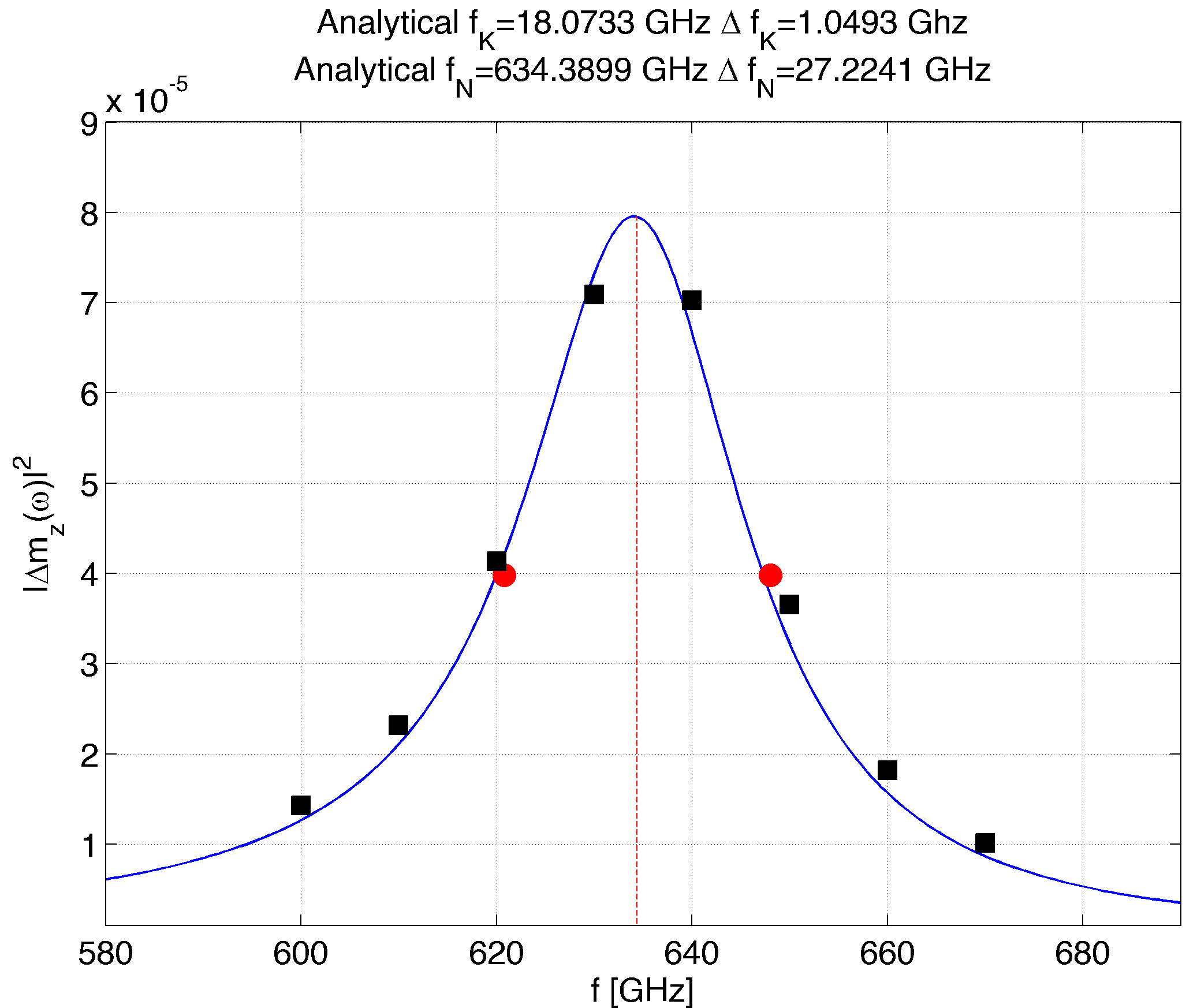}
    \caption{Frequency response power spectrum $|\Delta \tilde{m}_z(\omega)|^2$. detail of the nutation peak. The FMR frequency is around 18 GHz whereas the nutation frequency is about 634 GHz. The respective linewidths are about 1 GHz and 27 GHz. Blue line is eq.\eqref{eq:iLLG an freq resp}, red dots and dashed line are analytical formulas \eqref{eq:resonance frequencies},\eqref{eq:resonance linewidths}, black symbols are the result of numerical simulations of iLLG dynamics with IMR-MS1.}
    \label{fig:freq resp}
\end{figure}

Here we consider an infinite thin-film ($D_x=0, D_y=0, D_z=1)$ with material parameters: damping $\alpha=0.023$, saturation magnetization such that $\mu_0 M_s=0.93$T, inertial time scale $\tau=1.26$ ps.

In figure \ref{fig:lin iLLG time domain}, we report the time-evolution of $m_z$ during relaxation under zero bias field starting from an initial state tilted in the $x-z$ plane such that $m_x=0.01, m_z=0.01$. The analytical solution of eq.\eqref{eq:linear iLLG} is used to benchmark the proposed numerical techniques with different time step amplitudes. It is apparent that IMR technique allows the use of the largest time steps (up to $\Delta t=0.01$, around 25 samples per nutation period) yielding no significant loss of accuracy with respect to the analytical solution. One can also see that IMR-MS with first-order formula \eqref{eq:D2p3} (IMR-MS1) is accurate when $\Delta t\sim 0.0001$ (corresponding to 0.61 fs in physical units), while is not able to follow nutation dynamics after 5-6 periods (the period is 0.27) with 50 times larger time step $\Delta t=0.005$. Conversely, IMR-MS with second order formula \eqref{eq:D2p4} (IMR-MS2) performs well with $\Delta t=0.005$ (slightly above 50 samples per period) and provides a measured speedup of about 30 times compared to the former. This occurs since the average number of Newton iterations remains of the same order of magnitude, namely 2 for IMR-MS1 with $\Delta t=0.0001$ and 3 for IMR-MS2 with 50 times larger $\Delta t$ (the tolerance for Newton-Raphson iteration was set to $10^{-14}$). Finally, comparing IMR and IMR-MS2 methods, one can see that, despite correctly reproducing the nutation oscillation, IMR-MS2  produces a small phase-shift when the largest time step $\Delta t=0.01$ is chosen.

Next, by using eqs.\eqref{eq:linear iLLG}, the frequency response power spectrum of the out-of-plane magnetization component $m_z$ has been computed under a bias field $h_{ax}=0.35$T and ac field directed along $y$.
The iLLG equation \eqref{eq:iLLG} has been solved numerically in order to determine the frequency response power spectrum. Namely, given the susceptibility $\chi(\omega)$ in eq.\eqref{eq:iLLG an freq resp}
and the cross-power spectrum matrix $    (\Delta \tilde{\bm m})\cdot(\Delta \tilde{\bm m})^H=(\chi(\omega)\cdot  \tilde{\bm h}_\mathrm{ac})\cdot (\chi(\omega)\cdot  \tilde{\bm h}_\mathrm{ac})^H$ ($^H$ means conjugate transpose)
assuming that the only nonzero component of the input field $\tilde{\bm h}_\mathrm{ac}=(\tilde{h}_y,\tilde{h}_z)^T$ is $\tilde{h}_y=1$ (Fourier Transform of a Dirac delta) and the output response is the out-of-plane magnetization $\Delta \tilde{m}_z$, the output power spectrum is:
\begin{equation}
    |\Delta \tilde{m_z}(\omega)|^2=|\chi_{zy}(\omega)|^2  \quad,
\end{equation}
which is reported in fig.\ref{fig:freq resp} and compared with analytical formulas \eqref{eq:resonance frequencies},\eqref{eq:resonance linewidths}.
In order to compute $|\chi_{zy}|^2$ from time-domain numerical simulations, we apply a sinusoidal field $h_y(t)$ at frequency $\omega_0$ with sufficiently small amplitude (in order to stay in the linear regime) and measure the steady-state oscillation of $m_z(t)$ after sufficiently long time so that the transient response has vanished. This has been performed by choosing a simulated time $T=2$ns and ac field amplitude equal to 0.06T. 
Then, the Fast Fourier Transform $M_z(\omega)=\mathcal{F}[m_z(t)]$ has been computed and the maximum of the power spectrum $|M_z(\omega)|^2$ has been determined. This procedure has been repeated for several values of $\omega_0$. The so determined points form samples of the frequency response power spectrum $|\chi_{zy}(\omega)|^2$.

In figure \ref{fig:freq resp}, numerical simulations performed with IMR-MS1 with $\Delta t$ corresponding to 1 fs are used to compute the output power spectrum (black symbols). The results are in excellent agreement with analytical theory.

\subsection{Spatially-inhomogeneous spin nutation driven by terahertz applied field}

\begin{figure}[t]
    \centering
    \includegraphics[width=0.9\textwidth]{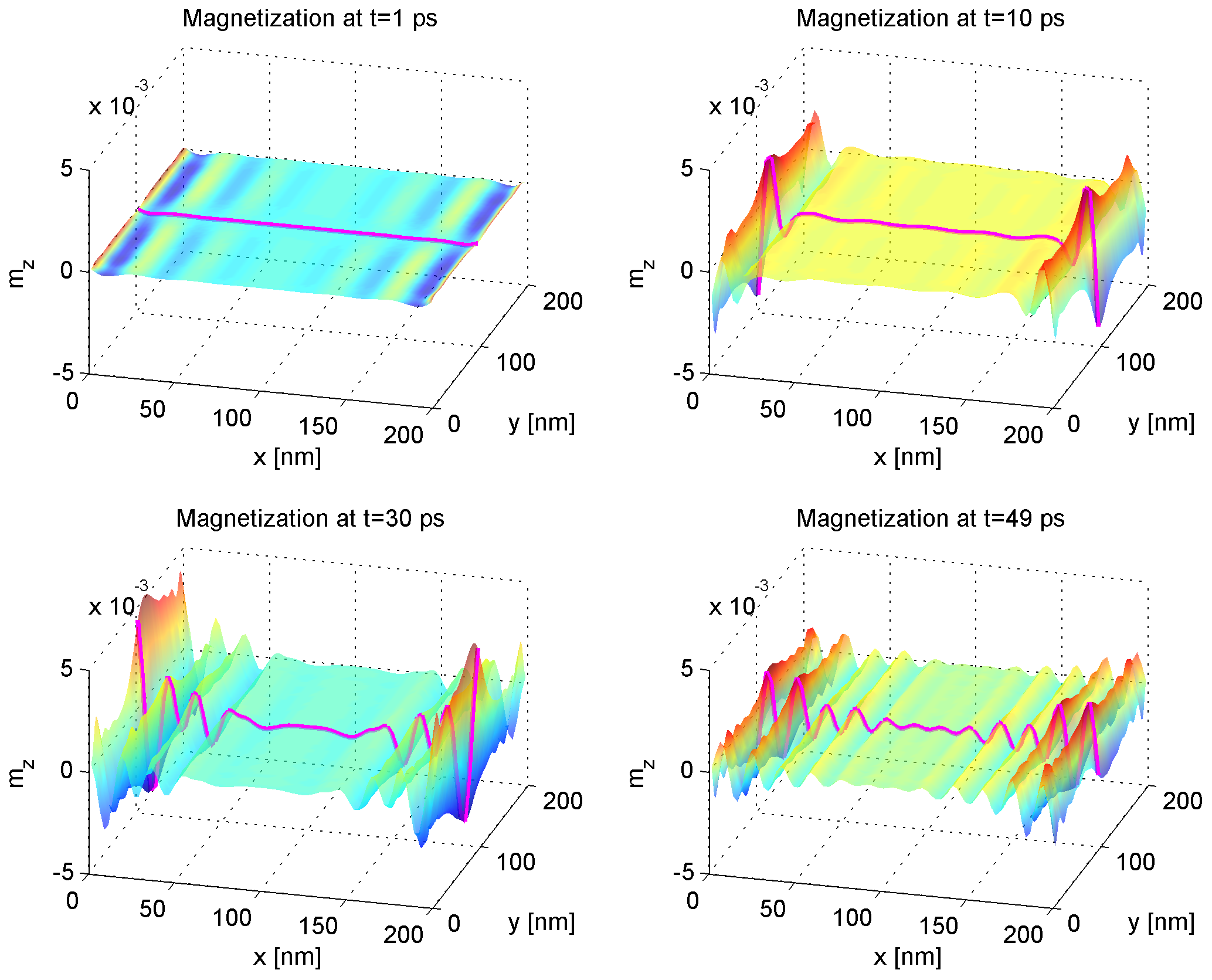}
    \caption{Spatial profiles of short-wavelength magnetization out-of-plane component $m_z$ obtained by FFT high-pass filtering at time $t=1,10,30,49\,$ps. The magenta solid line is a guide for the eye to follow the spin wave oscillation profile along the $x$ direction.}
    \label{fig:spin waves}
\end{figure}

In order to perform efficient time-domain micromagnetic simulations of iLLG dynamics with full spatial discretization, the proposed IMR-MS2 time-stepping has been implemented in the finite-difference numerical code MaGICo\cite{dAquino2005geometrical,MaGICo}, which retains the same computational cost as the simulation of classical precessional LLG dynamics while keeping important conservation properties as outlined above. 

To validate the code, we explore typical spatio-temporal patterns of iLLG dynamics considering the time-domain simulation of ultra-short inertial spin waves in a confined ferromagnetic thin-film. At terahertz frequencies, the behavior of small magnetization oscillations significantly deviates from the classical description of exchange-dominated spin-waves, in that an ultimate limiting propagation speed appears\cite{daquino2023micromagnetic}. This difference is mostly due to the mathematical structure of eq.\eqref{eq:iLLG} compared with the classical LLG precessional dynamics, i.e. the same equation where one sets $\xi=0$. In fact, when inertial effects are taken into account, the torque proportional to the second-order time-derivative transforms the classical LLG equation into a wave-like equation with hyperbolic mathematical character. In this respect, on short time scales, finite time delays are expected in magnetization response propagation far from local external excitation. 

The considered sample is made of Cobalt and has a thin-film shape with square cross-section $200\times 200\,$nm$^2$ and thickness $5\,$nm. The ferromagnetic nanodot is initially at equilibrium, being saturated along the $x$ axis by a static field $\mu_0 H_{ax}=100$ mT. 
The value of material parameters are $\gamma=2.211\times 10^5$ m$\,$A$^{-1}\,$s$^{-1}$, $\mu_0 M_s=1.6$ T, $A=13$ pJ/m ($l_\mathrm{ex}=3.57$ nm), $\tau=0.653$ ps $(\xi=0.0338)$ and $\alpha=0.005$.

The applied field is a spatially-uniform sine wave step (turned on at $t=0$) along the $y$ axis transverse to the equilibrium configuration with amplitude $\mu_0 H_{ay}=100$ mT and frequency $f=1386$ GHz. For the above choice of parameters, this excitation frequency is slightly above the nutation resonance frequency and corresponds to inertial spin waves with wavelength around 20 nanometers\cite{daquino2023micromagnetic}.  

The numerical simulation of iLLG equation \eqref{eq:iLLG} is performed with a time-step of 25 fs, a $2.5\times 2.5\times 5$ nm$^3$ computational cell, which corresponds to discretize the thin-film into $80\times 80$ square prims cell. In order to isolate short-wavelength spin wave propagation from the rest of the simulated spatial pattern, high-pass spatial filtering via two-dimensional Fast Fourier Transform is performed on magnetization components. The simulated time is $100\,$ps and some snapshots of the magnetization out-of-plane component $m_z$, taken at different time instants, are reported in figure \ref{fig:spin waves}. 

The magnetization is initially at the equilibrium and mostly aligned with the static field along the $x$ direction except close to the square corners where there is the most pronounced deviation. Such a tilting acts as local excitation for inertial spin waves when the time-varying ac field step is applied\cite{daquino2023micromagnetic}. In fact, as it can be seen in the various panels of fig.\ref{fig:spin waves}, two wavepacket with wavelength $\approx 20\,$nm propagate from the edges of the nanodot toward its center after the application of the ac field, consistently showing a propagation with finite speed $\approx 2000\,$m/s compatible with that predicted by the theory\cite{daquino2023micromagnetic}.   

\section{Conclusion}

In this paper, we have proposed second-order accurate and efficient numerical schemes for the time-integration of the ultra-fast inertial magnetization dynamics.  We have shown that the iLLG equation describes a higher-order dynamical system compared to the classical precessional dynamics which requires to double the degrees of freedom for its desription. We have derived the fundamental properties of the iLLG dynamics, namely conservation of magnetization amplitude and 'angular momentum' projection, Lyapunov structure and generalized free energy balance properties, and demonstrated that the proposed implicit midpoint rule (IMR) time-stepping is able to correctly reproduce them unconditionally. Suitable Newton technique has been developed for the inversion of the nonlinearly coupled system of equations to be solved at each time-step. For large-scale micromagnetic simulations with full spatial discretization, efficient numerical time-stepping schemes based on implicit midpoint rule combined with appropriate multi-step method for the inertial term, termed IMR-MS of order 1 and 2, have been proposed. These schemes retain the same computational cost of the IMR for the classical LLG dynamics while providing conservation of magnetization amplitude and accurate reproduction of the high frequency nutation oscillations. In particular, thanks to the unconditional stability due to its implicit nature along with the second-order accuracy on the inertial term, both the IMR and IMR-MS2 allow choosing  moderately large time-steps only based on accuracy requirements for the description of the nutation dynamics. The proposed techniques have been successfully validated against test cases of spatially-homogeneous and inhomogeneous magnetization iLLG dynamics demonstrating their effectiveness. For these reasons, we believe that these numerical schemes can become a standard de facto in the micromagnetic simulation of inertial magnetization dynamics in nano- and micro-scale magnetic systems.

\section*{Acknowledgements}
M.d'A., S.P. and C.S. acknowledge support from the Italian Ministry of University and Research, PRIN2020 funding program, grant number 2020PY8KTC.

\bibliographystyle{elsarticle-num}
\bibliography{main}

\end{document}